\documentstyle[11pt,newpasp,twoside,epsf]{article}
\def\edcomment#1{\iffalse\marginpar{\raggedright\sl#1\/}\else\relax\fi}
\reversemarginpar

\begin{document}
\title{Cosmic Data Fusion}
\author{S. L. Bridle}
\affil{
OMP, 
14 Av. E. Belin, 31400 Toulouse, France;
Cavendish Laboratory,  
Madingley Road, 
Cambridge CB3 0HE, UK
}

\begin{abstract}
We compare and combine likelihood functions of the cosmological
parameters $\Omega_{\rm m}$, $h$ and $\sigma_8$ from the CMB, 
type Ia supernovae and from probes of large scale structure.
We include
the recent results from the CMB experiments BOOMERANG and
MAXIMA-1. Our analysis assumes a flat $\Lambda$CDM cosmology with a
scale-invariant adiabatic initial power spectrum.
First we consider three data sets that
directly probe the mass in the Universe, without the need
to relate the galaxy distribution to the underlying mass
via a ``biasing'' relation: peculiar velocities, CMB and supernovae.
We assume a baryonic fraction as inferred from Big-Bang Nucleosynthesis
and find that all three data sets agree well,
overlapping significantly at the 2$\sigma$ level. This therefore
justifies a joint analysis, in which we find a joint best
fit point and $95\%$ confidence limits of $\Omega_{\rm m}=0.28\,
(0.17,0.39)$, $h=0.74\, (0.64,0.86)$, and $\sigma_8=1.17\,
(0.98,1.37)$. 
Secondly we extend our earlier work on combining CMB, supernovae,
cluster number counts, IRAS galaxy redshift survey data 
to include BOOMERANG and MAXIMA-1 data and to allow a free 
$\Omega_{\rm b} h^2$. We find that, given our assumption of a scale
invariant initial power spectrum ($n=1$), we obtain the robust result
of $\Omega_{\rm b} h^2= 0.031 \pm 0.03$, which is dominated by the
CMB constraint.
\end{abstract}

\section{Introduction}

A simultaneous analysis of the constraints placed on the cosmological
parameters by various different kinds of data is essential because
each different probe typically constrains a different combination of
the parameters. By considering these constraints together, one can
overcome any intrinsic degeneracies to estimate each fundamental 
parameter and its corresponding random uncertainty. The comparison 
of constraints can also provide a test for the validity of the 
assumed cosmological model or, alternatively, a revised evaluation 
of the systematic errors in one or all of the data sets. Recent 
papers that combine information from several data sets 
simultaneously include Gawiser \& Silk (1998), Bridle et al. (1999; 2000),
Bahcall et al. (1999), Bond \& Jaffe (1999), Lineweaver (1998) and 
Lange et al. (2000).

The anisotropies in the Cosmic Microwave Background (CMB) depend on
the state of the universe at the epoch of recombination, on the 
global geometry of space-time and on any re-ionization. Thus they 
provide a powerful and potentially accurate probe of the 
cosmological parameters (see Hu, Sugiyama and Silk 1997 for a 
review). With the recent release of results from a new generation of
CMB experiments BOOMERANG and MAXIMA-1 have come a number of 
parameter estimation analyses, including those by Lange et al. 
(2000), Balbi et al. (2000) and Tegmark \& Zaldarriaga (2000).
Constraints from the CMB are discussed in Section 2.

Galaxy motions relative to the Hubble flow arise from the 
gravitational forces due to mass-density fluctuations; they 
therefore reflect the underlying distribution of matter (both dark 
and luminous), and can thus provide constraints on the cosmological 
density parameter $\Omega_{\rm m}$ and the fluctuation amplitude 
parameter $\sigma_8$. 
The distances of type Ia supernovae (SN) can now be measured at 
large redshift. Thus they can provide constraints on the 
acceleration of the universal expansion, and the corresponding 
parameters $\Omega_{\rm m}$ and $\Omega_{\Lambda}$, via a classical 
cosmological test based on the luminosity-redshift relation. 
Velocities, supernovae and CMB
allow direct dynamical constraints free 
of assumptions regarding the ``biasing" relation between the 
distribution of galaxies and the underlying matter density, which 
are unavoidable when interpreting galaxy redshift surveys. 
Cosmological parameter estimates are presented in Section 3
(for more details on this work see Bridle et al. 2000).

In Bridle et al. (1999) we investigated the combination of
constraints from CMB data, the abundance of clusters of galaxies 
(Eke et al. 1998) and the IRAS 1.2 Jy redshift survey (Fisher, 
Scharf \& Lahav 1994). These data sets were found to be in excellent 
agreement. In Section 4 this work is updated to include the 
BOOMERANG-98 and MAXIMA-1 data, and also $\Omega_{\rm b} h^2$ is
included as a free parameter.

\section{CMB constraints}

\begin{figure}
\plotfiddle{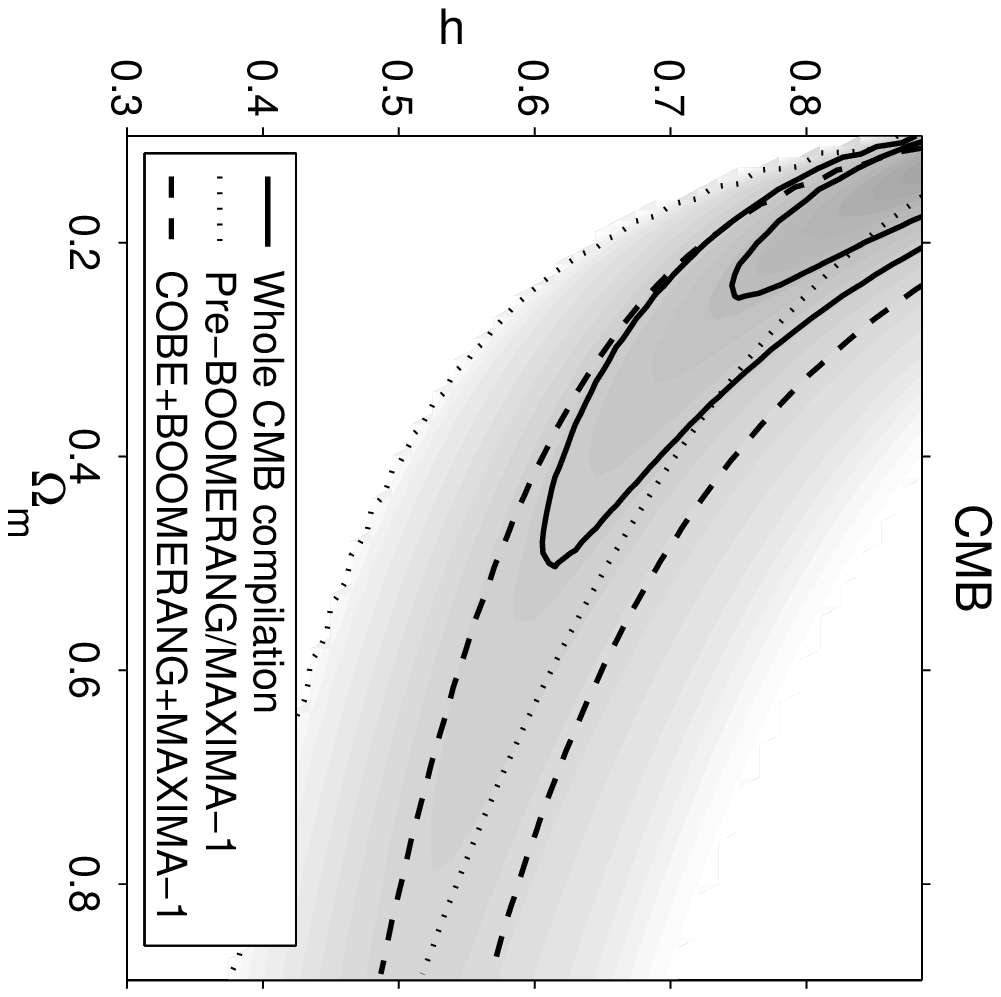}{7cm}{90}{85}{85}{350}{-130}
\plotfiddle{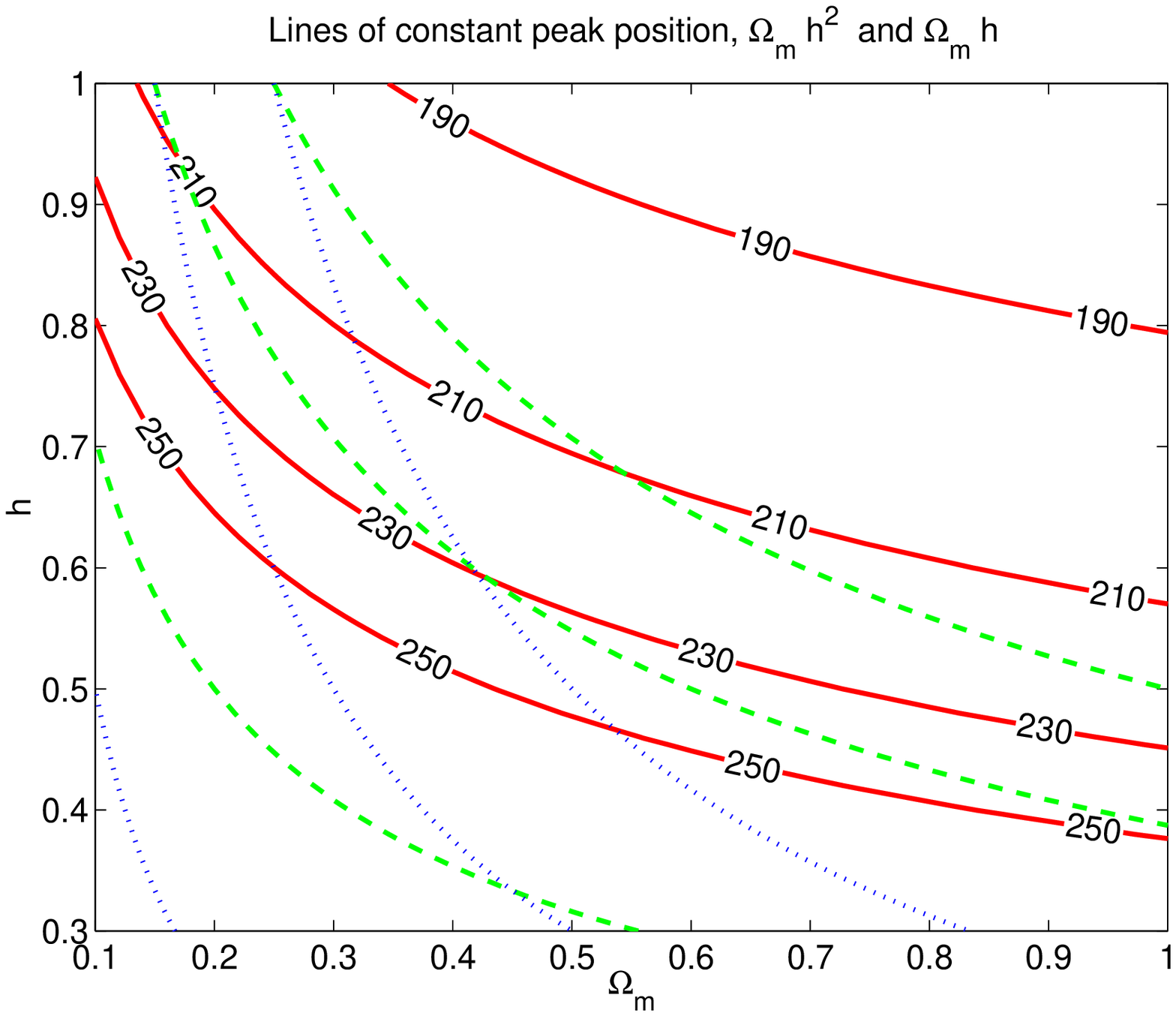}{7cm}{0}{50}{50}{-150}{-100}
\caption{
Top: Constraints in the $\Omega_{\rm m}$, $h$ plane from the CMB 
data (marginalised over $\sigma_8$ and with $\Omega_{\rm b} h^2$ fixed
at $0.019$). Solid lines show the $68$, $95$ \%
limits using the whole CMB data compilation. 
$95$ \% contours from the pre-BM and from just 
COBE$+$BOOMERANG-98$+$MAXIMA-1 data are shown by the dotted and 
dashed lines respectively. \\
Bottom:
Dashed lines are at $\Omega_{\rm m} h^2 = 0.05, 0.15, 0.25$.
Dotted lines are at $\Omega_{\rm m} h = 0.1, 0.2, 0.3$.
Solid lines are at the peak positions labelled, given
$\Omega_{\rm m}+\Omega_{\Lambda}=1$, $\Omega_{\rm b}h^2=0.019$.
}
\end{figure}

We use the same compilation of CMB anisotropy measurements as in 
Bridle et al. (2000), marginalising
over the $10$ and $4$ per cent calibration 
uncertainties quoted, respectively, for the BOOMERANG-98
and MAXIMA-1 results, fully taking into account the correlated 
nature of the calibration errors (Bridle et al., in preparation).
We assume that the universe is flat ($\Omega_{\rm m} +\Omega_{\Lambda}=1$) 
with a scale invariant initial power spectrum ($n=1$), with negligible
tensor contrubutions and negligible re-ionization. We obtain
theoretical CMB power spectra as a function of the cosmological
parameters using the CMBFAST and CAMB codes (Seljak \& Zaldarriaga, 
1996; Lewis, Challinor \& Lasenby, 2000).

The COBE data constrain the large scale temperature fluctuations 
well, which converts to a strong constraint on $\sigma_8$ for given 
values of $h$ and $\Omega_{\rm m}$. The CMB data indicate the 
position of the first acoustic peak, near $\ell\sim 200$ which 
corresponds to a wavenumber of 
$k \sim 0.03\,\, h \,\,{\rm Mpc}^{-1}$. This constrains the 
combination $\Omega_{\rm m}+\Omega_{\Lambda}$ to be roughly around 
unity (e.g. Efstathiou et al. 1999, Dodelson \& Knox 2000, Lange et 
al. 2000, Balbi et al. 2000, Tegmark and Zaldarriaga 2000), 
consistent with the flat universe assumed in our current analysis. 
In fact using just BOOMERANG and COBE, Lange et al. (2000) find
$\Omega_{\rm m}+\Omega_{\Lambda} \sim 1.1$ (Fig.~2), whereas using 
just MAXIMA-1 and COBE, Balbi et al. (2000) find $\Omega_{\rm m}+
\Omega_{\Lambda} \sim 0.9$. 

At $\sim 1^\circ$ angular scales the 
height of the first acoustic peak constrains the matter-radiation
ratio at last scattering, and this ratio is proportional to $\Omega_{\rm m}
h^2$. In addition, given our assumption of a flat universe, 
$\Omega_{\rm m}$ and $h$ also significantly affect the position of 
the first acoustic peak (see Fig 2. of White, Scott and Pierpaoli 
2000 for an illustration). Increasing $\Omega_{\rm m}$ moves the 
peak to lower $\ell$, as does increasing $h$. These two effects 
combine to give the likelihood distribution in the 
$\Omega_{\rm m}$-$h$ plane shown in the left hand panel of Fig.~1 
(for $\Omega_{\rm b} h^2=0.019$). The 
slightly lower first peak height indicated by the BOOMERANG and 
MAXIMA-1 data and the lower $\ell$ position of the first peak from 
the BOOMERANG data produce a constraint at higher $\Omega_{\rm m}$ 
and $h$ than does the pre-BOOMERANG/MAXIMA-1 compilation (hereafter 
pre-BM). Using the whole compilation together defines a region in 
$(\Omega_{\rm m}$,$h)$ space at the intersection of the 
BOOMERANG$+$MAXIMA-1 and the pre-BM contours. This occurs at high 
$h$ and low $\Omega_{\rm m}$. It is interesting to note that the 
degeneracy directions for each of the pre-BM and the 
BOOMERANG$+$MAXIMA-1 data sets are somewhat different (as shown in 
Fig. 1b.). One possible explanation for this is that the older data 
put a strong constraint on the peak height, which is a function of 
$\Omega_{\rm m} h^2$. On the other hand the BOOMERANG$+$MAXIMA-1 
data, with their detailed $\ell$ space coverage but significant 
calibration uncertainties, place a strong constraint on the peak 
position. Lines of constant peak position lie more parallel to the 
$\Omega_{\rm m}$ axis than do lines of constant 
$\Omega_{\rm m} h^2$, shown in Fig.~1 
(derived from Efstathiou and Bond 1999, discussed in more detail in 
Bridle 2000). Therefore using the whole CMB compilation
allows tighter constraints to be placed on $h$ and 
$\Omega_{\rm m} $. 

Increasing $\Omega_{\rm b} h^2$ (for fixed $\Omega_{\rm m} h^2$)
incraeses the first peak height while decreasing the second peak 
height, which improves the fit to the BOOMERANG-98 and MAXIMA-1 data,
as discussed elsewhere in these proceedings.
If $\Omega_{\rm b} h^2$ is changed to 0.03 from 0.019, the preferred
region in the $\Omega_{\rm m}$, $h$ plane (Fig.~1) shifts to higher
$\Omega_{\rm m}$ and $h$ (since increasing $\Omega_{\rm m} h^2$ 
brings the first peak height back down).

\section{Comparison and Combination with Peculiar Velocities and Supernovae}

We use the SFI peculiar velocity catalogue (Haynes et al. 
1999a,b) which consists of $\sim\!1300$ spiral galaxies.
The analysis follows in general the maximum-likelihood method of 
Zaroubi et al. (1997) and Freudling et al. (1999). 
Note that the linear analysis of the velocity data addresses the 
scaled  power spectrum $P(k) \Omega_{\rm m}^{1.2}$ rather than 
$P(k)$ itself, and it therefore constrains the combination of 
parameters $\sigma_8 \Omega_{\rm m}^{0.6}$, which serves as a 
measure of the power-spectrum amplitude. 
This result is almost
independent of $\Omega_{\Lambda}$ (Lahav et al. 1991).
In order to account for nonlinear effects acting on small scales, we 
add to the linear velocity correlation model an additional free 
parameter, $\sigma_{\rm v}$, representing an uncorrelated velocity 
dispersion at zero lag (this is discussed in more detail in 
Silberman et al., in preparation, and Bridle et al. 2000).
In the absence 
of any other information, we have carried out the Bayesian 
procedure for the case where we have no knowledge of a free 
parameter: we have marginalised over $\sigma_{\rm v}$.
The velocity data constraints at the $95\%$ confidence level are
$0.48<\sigma_8 \Omega_{\rm m}^{0.6} <0.86$ and $0.16<\Omega_{\rm m} 
h< 0.58$, with roughly uncorrelated errors. 

We use the supernova Ia 
constraints obtained by Perlmutter et al. (1999), which 
are fully consistent with those of Riess et al. (1998), based on 
applying the classical luminosity-redshift test to distant type Ia
supernovae. The sample consists of 42 high-redshift SN ($0.18 \leq
z\leq0.83$), supplemented by 18 low-redshift SNe ($z < 0.1$). This
analysis determines a combination of $\Omega_{\rm m}$ and
$\Omega_{\Lambda}$. Note that, unlike PV and CMB, SN are 
insensitive to the form of the matter power spectrum and depend 
only on the overall geometry of the universe. Since we limit 
ourselves in this paper to a flat universe, the SN constraint is 
translated to a likelihood function of $\Omega_{\rm m}$,
roughly $0.13 < \Omega_{\rm m} < 0.44$.

\begin{figure}
\plotfiddle{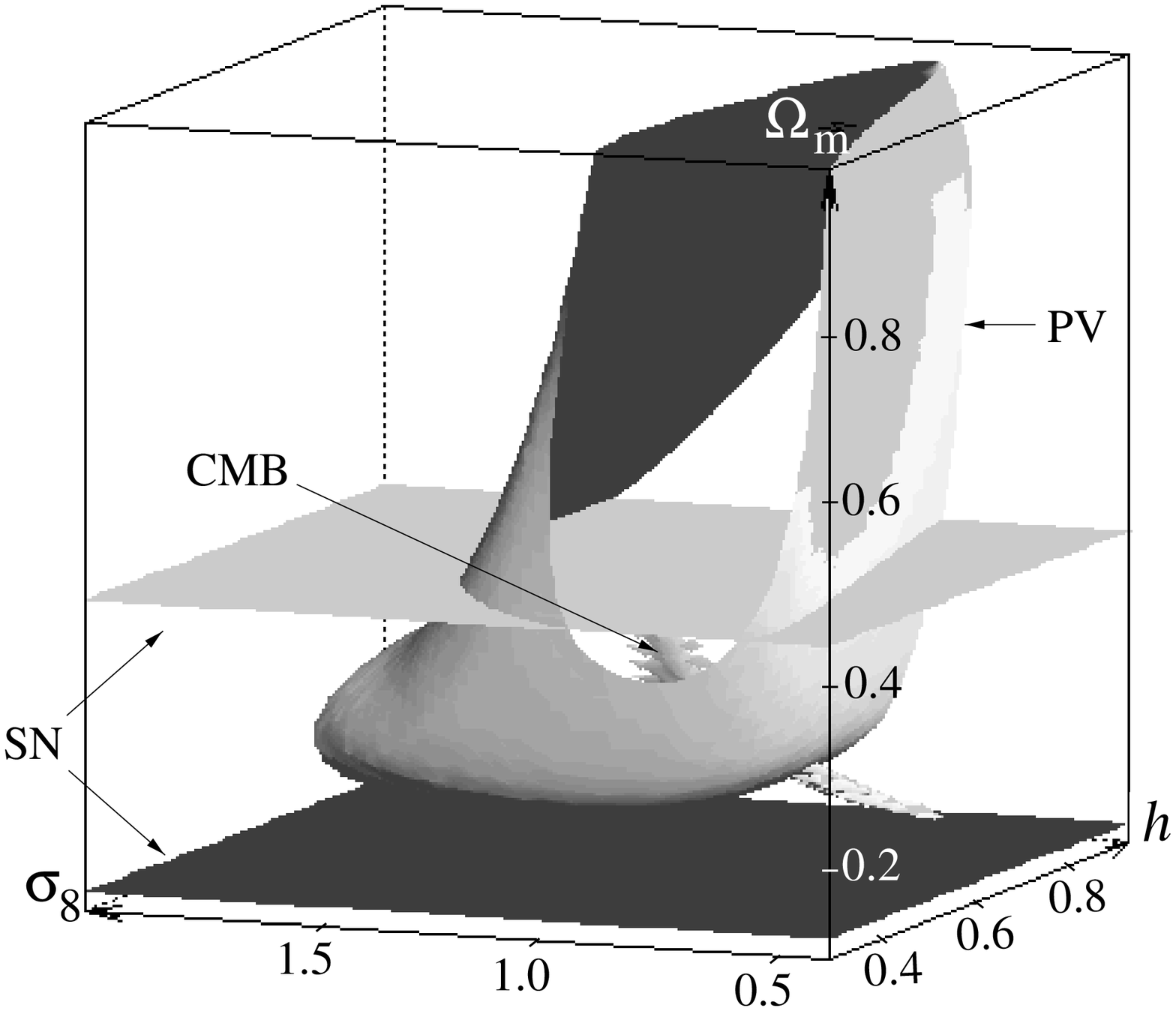}{8cm}{0}{45}{45}{-130}{0}
\plotfiddle{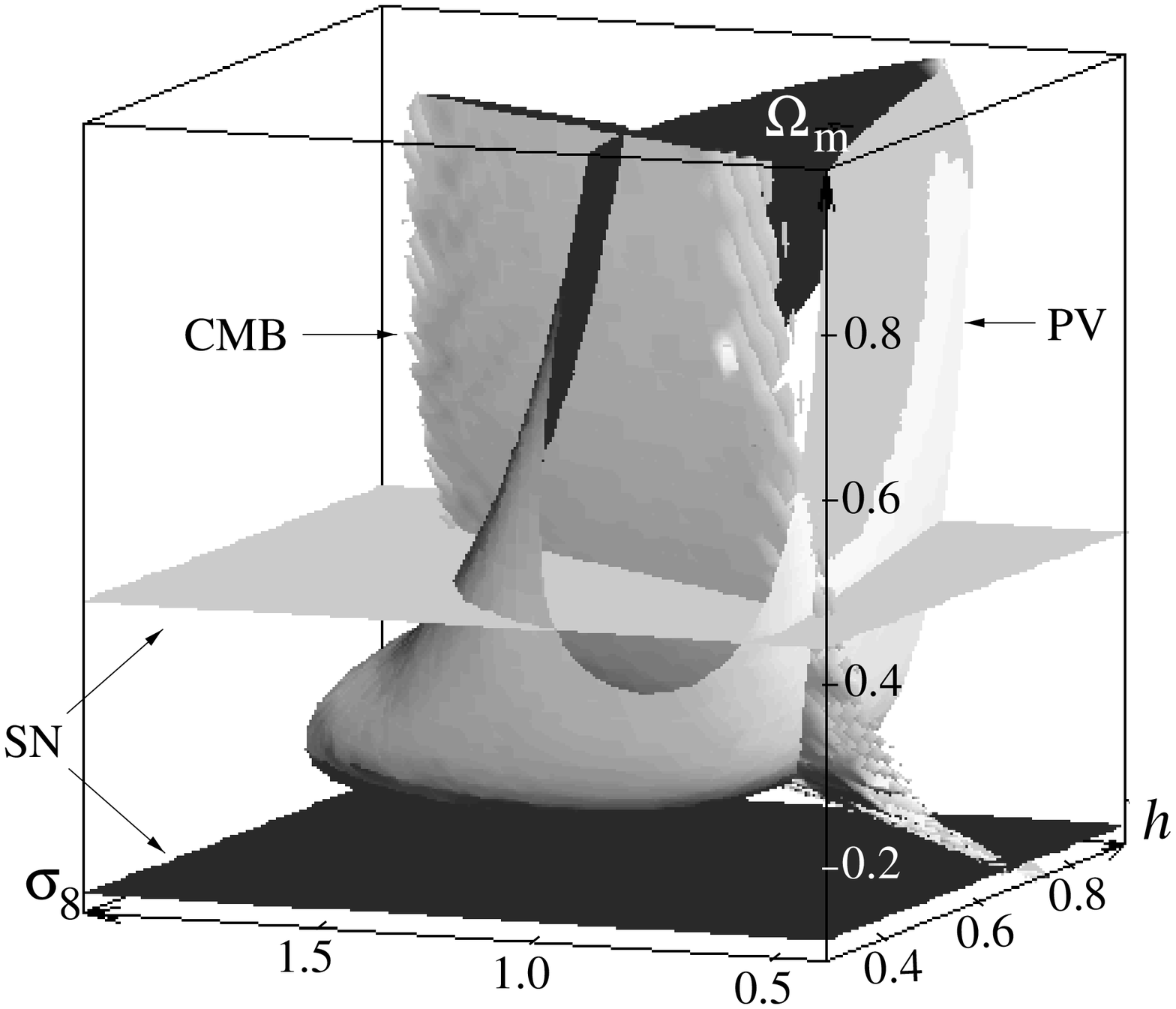}{8cm}{0}{45}{45}{-130}{0}
\caption{
Top: PV, CMB (whole compilation) and SN 2$\sigma$ 
iso-probability surfaces. For PV and CMB the surfaces are at 
$\Delta$log(Likelihood)=4.01, and for the SN the surfaces are at 
$\Delta$log(Likelihood)=2.00, corresponding to the 95 per cent 
limits for 3 and 1 dimensional Gaussian distributions 
respectively. 
The SN surfaces are two 
horizontal planes. 
Bottom: the same but this time the data used for the CMB 
surface is the pre-BM data.
}
\end{figure}

In order to examine how well the constraints from PV, CMB and SN 
agree with each other we plot in Fig.~2 the three 
corresponding iso-likelihood surfaces, at the 2-sigma level, in the 
three-dimensional parameter space $(h,\sigma_8,\Omega_{\rm m})$. 
The upper and lower $95$ per cent limits on $\Omega_{\rm m}$ from 
SN are the two horizontal planes. The PV surface encloses a space 
at roughly constant $\Omega_{\rm m} h$ and $\sigma_8 
\Omega_{\rm m}^{0.6}$. The CMB surface lies in the intersection of 
the regions allowed by each of SN and PV. The fact that the 
constraints have a common region of overlap is not trivial; it 
indicates a reasonable goodness of fit between the three data sets 
within the framework of the assumed cosmological model, which 
justifies a joint likelihood analysis aimed at parameter 
estimation. To illustrate the complementary nature of these three 
data sets and
for comparison the result of using the pre-BM CMB data instead is 
shown in the bottom left panel of Fig.~2.
 
The best fit cosmological parameters
given all three data sets are 
$h=0.74$, $\Omega_{\rm m}=0.28$ and $\sigma_8=1.17$,
from 
which we can derive $\sigma_8\Omega_{\rm m}^{0.6}=0.54$, 
$\Omega_{\rm m} h = 0.21$, $Q_{\rm{rms-ps}}=19.7 \mu$K and the age 
of the universe is $13.2$ Gyr. 
We may evaluate the probability of a single cosmological parameter,
independent of the values of the other cosmological parameters, by
integrating the probability over the values of the other 
parameters. This is what we mean by `marginalisation'. The solid 
lines in Fig.~3 shows the resulting 1-dimensional 
marginalised likelihood distributions for each parameter. We obtain 
the $95$ per cent limits by integrating the one- dimensional 
likelihood distributions and requiring that $95$ per cent of the 
probability lies between the quoted limits: 
$0.64 < h              < 0.86$,
$0.17 < \Omega_{\rm m} < 0.39$,
$0.98 < \sigma_8       < 1.37$
The $h$ range agrees well with that from the 
HST key project of $h=0.72 \pm 0.08$ (1$-\sigma$, Freedman et al. 2000) 
and the
$\Omega_{\rm m}$ limits are roughly centered on the popular value 
of $0.3$. 
\begin{figure}
\plotfiddle{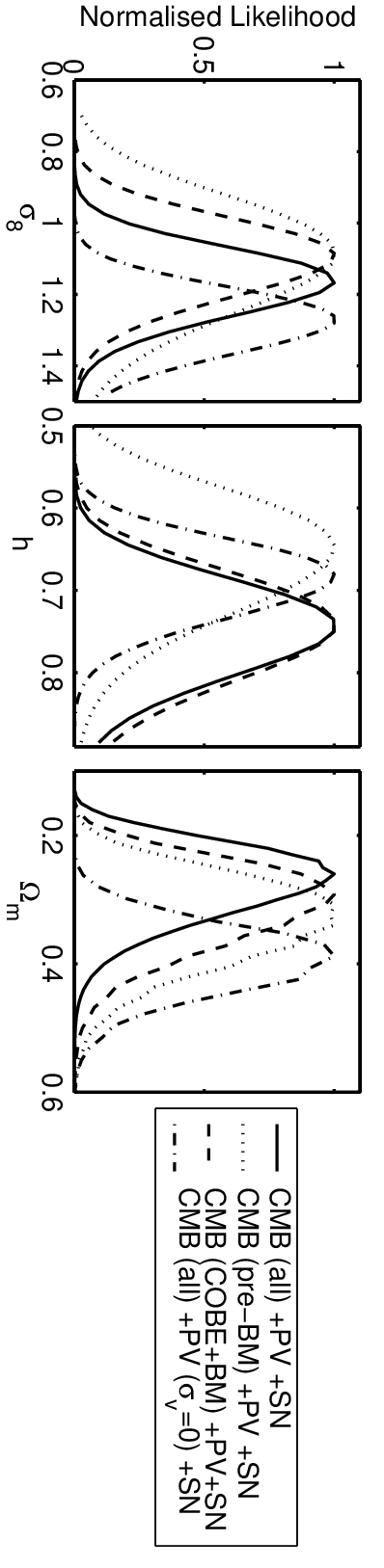}{4cm}{90}{55}{55}{250}{-120}
\caption
{The 1-dimensional marginalised likelihood distributions from the 
joint PV, CMB and SN likelihood function. Our main results are 
shown by the solid lines, which use the whole CMB data compilation, 
PV (marginalised over $\sigma_{\rm v}$) and SN. The dotted lines 
show the likelihood functions when PV, SN and just the pre-BM CMB 
data are used. The dashed line is the result when the CMB data is
just COBE + BOOMERANG + MAXIMA-1. The result (using all the CMB 
data) when the uncorrelated velocity dispersion term is not 
included in the PV analysis ($\sigma_{\rm v}=0$) is shown by the 
dot-dashed line.
}
\end{figure}

We have repeated the entire analysis using different subsets of CMB
data. Using the pre-BM data the 1-dimensional marginalised 
likelihood functions (dotted lines in Fig.~3) are in good 
agreement but somewhat wider than when using all the data, 
especially in the constraint on $h$ which extends to lower values 
than before. Using just the COBE, BOOMERANG-98 and 
MAXIMA-1 data (dashed lines) the results are very similar to when all
data is used. At first this may seem surprising given the much 
larger 3-dimensional surface, but the high 
$\Omega_{\rm m}$, $h$ part is ruled out by both PV and SN, leaving 
virtually the same region as when all CMB data are used. 

In the region of the power spectrum where a second acoustic peak is
predicted, we note that our best fitting models are not a good fit 
to the data, producing more power than observed by both BOOMERANG 
and MAXIMA-1 (see for example the power spectra plotted in 
Bridle et al. 2000).

\section{Constraints on the Baryon Density}

In this section we allow the baryon density, $\Omega_{\rm b} h^2$,
to vary as a free parameter, along with $\Omega_{\rm m}$, $h$ and
$\sigma_8$. We investigate combining the CMB and supernova data used
above with 
the galaxy cluster number count data of Eke et al. (1998) and the
IRAS galaxy redshift survey spherical harmonic analysis of Fisher,
Scharf and Lahav (1994). 
These four data sets are found to enclose a common volume in 
the four dimensional parameter space and thus a joint analysis 
is reasonable. The resulting one-dimensional marginalised constraints
are shown in Fig.~4. The $h$ value is similar to in the previous
section. The $\sigma_8$ value is somewhat lower than before, reflecting
the lower $\sigma_8 \Omega_{\rm m}^{0.6}$ value preferred by cluster
number counts compared to that found from velocities.
The $\Omega_{\rm m}$ value is now lower than before, at around $0.2$,
which is suprising given our comments at the end of Section 2. 
However a detailed examination in the four dimensional space reveals
that this is due to the precise way in which the cluster number count
constraint intersects with the CMB constraint at these high $h$ values.
The $\Omega_{\rm b} h^2$ value is similar to that found by 
Jaffe et al. (2000) (see also Bond in these proceedings), but because
of our assumption of $n=1$, the error bars are much smaller, ruling
out the nucleosynthesis value of $\Omega_{\rm b} h^2=0.019$ 
(Burles et al. 1999) at the
3 to 4 sigma level.
On using various subsets of the data, the constraints on all parameters
but  $\Omega_{\rm b} h^2$ vary by around 1 sigma, but this 
$\Omega_{\rm b} h^2$ result remains robust.
This is due to the fact that all data sets apart from the CMB are 
relatively insensitive to the value of $\Omega_{\rm b} h^2$.
Also the $\Omega_{\rm b} h^2$ value is not too strongly coupled to the
other cosmological parameters, relative to the tightness of the CMB 
constraint.

\begin{figure}
\plotfiddle{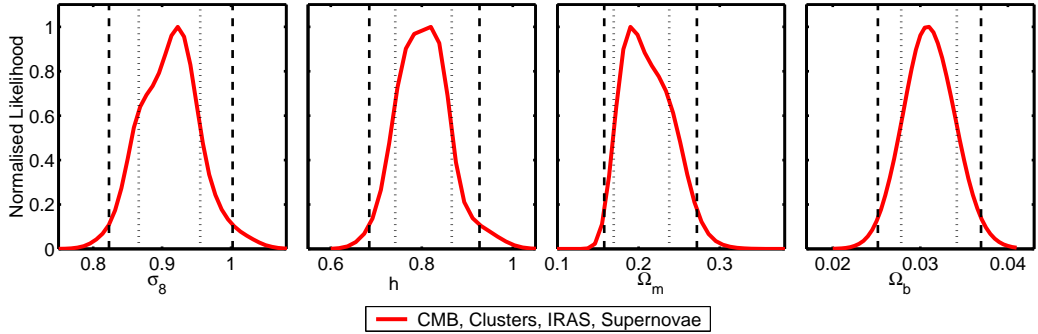}{6cm}{90}{50}{50}{200}{-100}
\caption
{The 1-dimensional marginalised likelihood distributions from the 
joint CMB, SN, cluster and IRAS likelihood function. 
}
\end{figure}

\section{Conclusion}

The addition of BOOMERANG and MAXIMA-1 to our CMB data compilation
brought down the height of the first acoustic peak and shifted it to 
larger angular scales, which both increase a combination of $h$ and 
$\Omega_{\rm m}$. The combination of BOOMERANG and MAXIMA-1 with 
the older CMB data had the effect of breaking the degeneracy 
between $h$ and $\Omega_m$ and leaving a high $h$ region of 
parameter space. 

We have performed a joint analysis of three complementary data sets
free of galaxy-density biasing, using peculiar velocities, CMB 
anisotropies, and high-redshift supernovae. The constraints from the 
three data sets overlap well at the 2-sigma level and there is  
acceptable goodness of fit. These data sets constrain roughly 
orthogonal combinations of the cosmological parameters, and are 
combined to provide tighter constraints on the parameters.
The values obtained from the joint analysis for $h$ and 
$\Omega_{\rm m}$, and for the combinations of cosmological 
parameters, are in general agreement with other estimates 
(eg. Bahcall et al. 1999), but this analysis tends to favor a 
slightly higher value for $\sigma_8$. 
The resulting constraint on the Hubble constant, 
$h=0.75 \pm 0.11$ ($95$ per cent confidence), agrees well with that 
from the HST key project value of $h=0.72 \pm 0.08$. This result is 
also similar to that of Lange et al. (2000, Table~1, P10). 

Motivated by the low second peak height implied by the BOOMERANG-98 and
MAXIMA-1 results, we consider the effect of including $\Omega_{\rm b} h^2$
as a free parameter. Our result for $\Omega_{\rm b} h^2$ is robust
at a similar value to that found by other authors (including
Jaffe et al. 2000 and Bond in these proceedings), but because we
have assumed a scale invariant initial power spectrum, our error
bars are approximately half the size.

Note that in this analysis we take all the data sets used at equal
weight. An extension to this work would be to allow freedom in the
weights given to the different probes, as discussed by Lahav in this
proceedings and in Lahav et al. (2000).

\acknowledgements

I thank my collaborators, O. Lahav, A. Lasenby, M. Hobson, I. Zehavi,
A. Dekel, C. Frenk, V. Eke, J. P. Henry, S. Cole for their contribution.
I also thank G. Rocha for her work on compilation of the CMB data
set, G. Efstathiou for providing the supernova likelihoods and
the referee of work included here, S. Zaroubi.

\bibliographystyle{/opt/TeX/tex/bib/mn}

\label{lastpage}

\end{document}